\documentclass[twocolumn,prx,aps,superscriptaddress]{revtex4}
\usepackage{amsmath,amssymb}
\usepackage[colorlinks, citecolor=blue]{hyperref}
\usepackage{amsfonts}
\usepackage{mathrsfs}
\usepackage{graphicx}
\usepackage{amssymb}
\usepackage{color}
\usepackage{dcolumn}
\usepackage{bm}
\usepackage{tikz,pstricks}
\usepackage[header,title,page,titletoc]{appendix}
\usepackage{url}

\begin{document}

\title{Fermionic criticality with enlarged fluctuations in Dirac semimetals}
\author{Jiang Zhou}
\affiliation{Department of Physics, Guizhou University, Guiyang 550025, PR China}
\author{Su-Peng Kou}
\thanks{Corresponding author}
\email{spkou@bnu.edu.cn}
\affiliation{Center for Advanced Quantum Studies, Department of Physics, Beijing Normal
University, Beijing 100875, China}

\begin{abstract}
The fluctuations-driven continuous quantum criticality has sparked
tremendous interest in condensed matter physics. It has been verified that
the gapless fermions fluctuations can change the nature of phase
transition at criticality. In this paper, we study the fermionic quantum
criticality with enlarged Ising$\times$Ising fluctuations in honeycomb
lattice materials.
The Gross-Neveu-Yukawa theory for the multicriticality between the semimetallic phase and two ordered
phases that break Ising symmetry is investigated by employing perturbative renormalization group approach.
We first determine the critical range in which the quantum fluctuations may render the phase transition continuous.
We find that the Ising criticality is continuous only when the flavor numbers of four-component Dirac fermions $N_f\geq1/4$.
Using the $\epsilon$ expansion in four space-time dimensions, we then study the Ising$\times$Ising multicriticality stemming from the
symmetry-breaking electronic instabilities. We analyze the underlying fixed-point structure and compute the critical exponents for the
Ising$\times$Ising Gross-Neveu-Yukawa universality class. Further, the correlation scaling behavior for the fermion bilinear on the honeycomb lattice at the multicritical point are also briefly discussed.
\end{abstract}

\maketitle

\section{Introduction}

The quantum phase transitions (continuous) at zero-temperature driven by
non-temperature parameters are believed to be key to understand some
unconventional properties of correlated many-body systems\cite{sachdev1,herbut1},
including strange metal phase in the high-temperature superconductors\cite{sachdev2,sachdev3,zgq},
ferromagnetic quantum criticality in the heavy fermion systems\cite{yhq} and deconfined quantum critical point (QCP)\cite{senthil1,senthil2}.
From a field-theoretical perspective, the general description for the continuous phase transition can be formulated in terms of the order parameter which acquires a nonzero value as the system is tuned across the transition.
Together with the renormalization group (RG) theory\cite{wilson1},
the Landau-Ginzburg-Wilson (LGW) paradigm provides a well understanding of the universal criticality and a effective method to calculate the critical exponents near the critical point, for example the extensively studied scalar $O(N)$ model which captures the continuous critical behavior of a wide variety of systems \cite{herbut1,kos1,zerf1}.

However, the LGW paradigm has been challenged recently by various examples in which the fluctuations from emergent degrees of freedom render the transition continuous.
The prime example is the transition between Neel and valence bond solid phase that separated by the deconfined QCP on the 2D square Heisenberg antiferromanets\cite{senthil1,senthil2}. A large number of theoretical and numerical studies demonstrate that the deconfined QCP is continuous as a result of the emergent fractionalized "spinon" and noncompact U(1) gauge field\cite{sandvik1,shao1,lou1,nahum1,wang1}.
Since the new degrees of freedom such as spinons emerge right at the QCP,
the LGW theory is fail to describe the deconfined QCP purely in terms of the space-time fluctuations of order parameter.
Another example that goes beyond the LGW picture is the fermion induced quantum critical point (FIQCP) which has attracted persistent attention in Dirac fermion systems
\cite{yao1,yao2,yao3,yao4,yin1,Scherer1,herbut3,herbut4,janssen1,
janssen2,classen1,classen2,classen3,roy1,roy2,roy3,roy4,lzx1,torres1,torres2,ihrig1,mihai1,liu1,ray1}.
Evidences for FIQCP have been embodied in the transition from semimetal to $\mathbb{Z}_3$ Kekule valence-bond-solid (VBS) phase of (2+1)D fermions on the honeycomb lattice\cite{yao1,yao2,classen1}. The Kekule VBS pattern breaks the translational symmetry and breaks the continuous
$U(1)$ symmetry down to $\mathbb{Z}_3$\cite{classen1,hou1,ryu1}, consequently, the Kekule phase transition allows a cubic term of VBS order parameter. From the view of point of LGW picture, the Kekule phase transition is expected to be first-order in the presence of cubic term of order parameter. Extensive studies, however, suggest that the presence of gapless fermion fluctuations at the critical point can dramatically change the nature of critical point and render a putatively first-order transition continuous\cite{yao1,classen1}. Various theoretical methods have been applied to study the FIQCP, ranging from perturbative RG \cite{yao3,ihrig1,mihai1,classen3}
to functional RG\cite{classen1,yin1,janssen1,knorr1}.
On the other hand, the sign-problem-free quantum Monte Carlo simulations for interacting fermions on lattice also push our
understanding of FIQCP \cite{wangl1,lzx2,chand1,wu1,wu2}.

Building on the concept of FIQCP, the quantum criticality in the presence of external fluctuations provided by the Dirac fermions
has sparked tremendous interest\cite{torres1,sorella1,otsuka1,tang1}.
A representative example is the semimetal-insulator transition of interacting electrons on the honeycomb lattice\cite{sorella1,tang1}.
When the interaction is sufficiently strong, the electrons may undergo continuous transition from semimetallic phase to a symmetry broken insulating phase characterized by an ordered ground state\cite{xu1}. A large on-site repulsive interaction, for example, gives rise to a spin-density-wave state\cite{knorr1,zerf2},
whereas a nearest-neighbor interaction would induce a charge-density-wave state\cite{ihrig1}.
A large number of interacting Dirac fermion model exhibit the continuous phase transition,
e.g., spin liquid\cite{boyack2}, superconducting and XY phases \cite{roy3,honerkamp1,boyack1,janssen,lang1},
and exotic topological phases\cite{vafek1,raghu1}.
Moreover, the recent theoretical studies suggest the multicritical point may be achieved in three ways:
(i) condensation of topological defects\cite{liuy1},
(ii) anticommuting mass terms\cite{sato1},
(iii) interaction instabilities\cite{Scherer1,classen3,roy3}.
It's argued that the quantum criticality occurs in the vicinity of these critical points and the universality class are defined through different microscopic models. Since the presence of external fluctuations provided by gapless Dirac fermions, these critical points are not captured by the conventional $O(N)$
universality classes. Instead, they are described by the Gross-Neveu-Yukawa theory, defining qualitatively different chiral Gross-Neveu-Yukawa (GNY) universality classes\cite{rosenstein1}.
Indeed, the critical behavior of a large number of transitions in condensed matter systems are captured by the GNY model.
To describe the critical behavior of GNY (or the gauged QED-GNY) theory, various theoretical methods and numerical methods have been applied, i.e., perturbative RG\cite{zerf1,ihrig1,yao2,mihai1,gracey1},
nonperturbative functional RG\cite{classen1,knorr1}, large-N method\cite{gracey2,zerf2} and quantum Monte Carlo simulations\cite{otsuka1,assaad1,meng1,meng2}.
In view of the symmetry of the broken phase, the GNY universality class comprises
chiral Ising ($\mathbb{Z}_2$) class \cite{rosenstein1,chand1,zerf1,mihai1,ihrig1},
chiral XY [O(2)] class\cite{yao1,janssen2,hands1,otsuka1,classen1,roy4}
and chiral Heisenberg[SU(2)] universality class\cite{knorr1,gracey2,zerf1,zerf2,lang1}.

Once the fermion-induced continuous criticality is established,
efforts to extend the FIQCP to the  multicritical point with enlarged fluctuations is significant,
as this theoretical problem relates to the competing orders and multicritical behavior of correlated electrons,
e.g., high-temperature superconductors\cite{sachdev2} and deconfined criticality\cite{senthil3,sato1}.
Recently, the studies of multicritical point are ongoing\cite{janssen2,torres2,classen3,jian1}.
It's demonstrated that the multicritical point is characterized by a emergent enlarged symmetry and features a continuous transition between two ordered phases as the system is tuned through the multicritical point, for example O(5) symmetry for O(3) Neel order and U(1) Kekule VBS.
Moreover, the multicritical point exhibits an enhanced symmetry within the Yukawa sector against small perturbations that break the O(5) symmetry\cite{torres2}.
These recent progress raises two fundamental issues: (1) How the multicritical behavior is modified by the interplay between two ordered phases and the gapless Dirac fermions. (2) To what extent does the multi-criticality affect the possible competing order parameters.

Here, we solve the two issues by exploring the gapless Dirac fermions coupled to the Ising$\times$Ising symmetry-breaking order parameters.
We first formulate the theory for the multicritical point with enlarged Ising$\times$Ising fluctuations, which we analysis using perturbative RG.
By including a cubic term in the theory of Ising FIQCP, we also identify the semimetal-CDW transition is continuous for the flavors of fermions $N_f$ fulfill $N_f>1/4$.
In particular, we find that the Kekule VBS phase can be enhanced by multicritical fluctuations, the crucial ingredient for the enhancement is the  anticommuting nature between the corresponding fermion bilinears and the Dirac gamma matrices in the kinetic part. Although our results are judged from the enlarged Ising$\times$Ising criticality, the enhancement scenario can be applied to other multicritical fluctuations, such as enlarged O(3)$\times$ U(1),  O(3)$\times$ Ising.

This paper is organized as follows. We formulate the critical theory with enlarged Ising$\times$Ising fluctuations in Sec. \ref{sec2}. After identifying the range in which the transition is continuous in Sec. \ref{sec3}, we perform RG analyses for the multicritical point in Sec. \ref{sec4}. We also show the Kekule VBS are enhanced by the multicritical fluctuations in Sec.\ref{sec4}. Conclusions are drawn in Sec. \ref{sec5}.

\section{Semimetal to Ising-order transition}\label{sec2}

We consider the spinless Dirac fermions on a honeycomb lattice, whose
low-energy effective theory in physical $2+1$ dimensions can be expressed as
the Lagrangian density\cite{classen1,classen3,roy3}
\begin{equation}
\mathcal{L}_{\psi }=i\bar{\Psi}\gamma ^{\mu }\partial _{\mu }\Psi ,
\end{equation}%
where the conjugate fermionic field $\bar{\Psi}=\Psi ^{\dag }\gamma ^{0}$,
the derivative operator reads $\partial _{\mu }=(\partial _{0},\partial
_{i}) $. Here the Fermi velocity $v_{F}=3t/2$ was set to unity for
convenience and the summation convention over repeated indices is assumed.
The $\gamma ^{\mu }$ matrices satisfy the Clifford algebra $\{\gamma ^{\mu
},\gamma ^{\nu }\}=2g^{\mu \nu }$, $\mu ,\nu =0,1,2,$ and $g^{\mu \nu }=$diag%
$(1,-1,-1)$ is a Minkowski space metric. We have defined the following $%
4\times 4$ Minkowski space gamma matrices
\begin{equation}\label{gamma}
\gamma ^{0}=\tau ^{0}\otimes \sigma ^{3},\gamma ^{1}=\tau ^{0}\otimes
i\sigma ^{1},\gamma ^{2}=\tau ^{3}\otimes i\sigma ^{2},
\end{equation}%
where the two-component identity matrix $\tau ^{0}$ and the standard Pauli
matrices $\tau ^{i}$ act on the valley indices ($K,-K$), the two-component
Pauli matrices $(\sigma ^{0},\sigma ^{i})$ act in sublattice space $(A,B)$.
In the free Dirac Lagrangian, the four-component Dirac spinor is defined as $
\Psi =(c_{AK},c_{BK},c_{A-K},c_{B-K})^{\text{T}}$. In the vicinity of Dirac
points, then the Bloch Hamiltonian reads $\mathcal{H}=\gamma ^{0}\gamma
^{i}k_{i}$ with the reduced Planck constant $\hbar =1$\cite{roy3}. There are two
matrices anticommute with all $\gamma ^{\mu }$ matrices
\begin{equation}
\gamma ^{3}=\tau ^{1}\otimes i\sigma ^{2},\gamma ^{5}=\tau ^{2}\otimes
i\sigma ^{2}.
\end{equation}%
We can define $\gamma ^{35}=i\gamma ^{3}\gamma ^{5}=\tau ^{3}\otimes i\sigma
^{0}$ which commutes with all $\gamma ^{\mu }$ but anticommutes with $\gamma
^{3}$ and $\gamma ^{5}$, or explicitly, $[\gamma ^{35},\gamma ^{\mu }]=0$, $
\{\gamma ^{35},\gamma ^{3}\}=0$, $\{\gamma ^{35},\gamma ^{5}\}=0$. It's
easily check that $[\gamma ^{35},\mathcal{H}]=0$. The Hamiotonian possesses
a symmetry implemented by $C\mathcal{H}C^{-1}=-\mathcal{H}$, where $C$ is
expressed as either $C=\gamma ^{0}$ or $C=\gamma ^{0}\gamma ^{35}$. This
symmetry is conventionally called chiral symmetry or sublattice symmetry on
bipartite graphene lattice. For generality, we introduce an arbitrary number
of $N_{f}$ fermion flavors of four-component Dirac fermions. The fermion
field carries a flavor index, $\Psi =\Psi _{i}$ with $i=\{1,2,...N_{f}\}$, $
N_{f}=2$ case corresponds to spin-1/2 fermions on the honeycomb lattice\cite{mihai1,ihrig1}.

We now turn to the instabilities accompanied by spontaneous breaking of the Ising (or $
Z_{2}$) symmetry. The patterns of $Z_{2}$ order-parameter can be achieved
through the condensation of Dirac fermion bilinears which gaps out the node
Dirac fermions. To this end, we introduce two mean field fermion bilinears: $
\langle \bar{\Psi}\Psi \rangle $ and $\langle \bar{\Psi}\gamma ^{35}\Psi
\rangle $, which can be triggered by sufficiently strong nearest-neighbor
electron interaction. Both patterns of fermion bilinears break the chiral
symmetry, the condensation of $\bar{\Psi}\Psi $ corresponds to
charge-density wave (CDW) and the condensation of $\bar{\Psi}\gamma
^{35}\Psi $ plays the role order parameter of quantum anomalous Hall (QAH)
phase.

We first consider the CDW phase. Introducing the $Z_{2}$ field $\chi
=\langle \bar{\Psi}\Psi \rangle $ which describes the fluctuating of CDW order
parameter, then the Lagrangian for the transition from the semimetal to CDW in
defined by the chiral Ising GNY model
\begin{equation}
\mathcal{L}_{\text{cICDW}}=\bar{\Psi}_{i}i\gamma ^{\mu }\partial _{\mu }\Psi
_{i}+g_{\chi }\bar{\Psi}\chi \Psi +\mathcal{L}_{\chi },
\end{equation}%
where the bosonic Lagrangian is given by
\begin{equation}
\mathcal{L}_{\chi }=\frac{1}{2}(\partial _{\mu }\chi )^{2}-\frac{1}{2}%
m_{\chi }^{2}\chi ^{2}-\lambda _{\chi }\chi ^{4}.
\end{equation}%
Here the parameters $m_{\chi }^{2}$ tunes the phase transition from
semimetallic phase to the phase with spontaneous $Z_{2}$ symmetry breaking
where the fermion mass are dynamically generated. To determine the nature of the transition, we introduce the cubic term in
the Landau-Ginzburg Lagrangian by hand, is of the form%
\begin{equation}
\mathcal{L}_{\text{cub.}}=b(\chi ^{3}+\chi ^{\ast 3}).
\end{equation}%
Such kind of terms also exist in the valence-bond-solid phase as the
redution of continuous symmetry down to discrete symmetry, i.e., $Z_{3}$ and
$Z_{4}$ symmetry\cite{Scherer1,classen1,yao2}. According to Landau criterion, the transition should be
first-order in the presence of cubic terms of order parameter in the
Lagrangian\cite{yin1}. In general space-time dimenisions $D$, the cubic coupling have
canonical dimensions $[b]=3-D/2$, which implies that the cubic terms is
strongly relevant near upper critical dimensions $D^{uc}=4$. By contrast,
the possible $Z_{4}$-anisotropy $\sim \chi ^{4}+\chi ^{\ast 4}$ on square
lattice is marginal and can be accessible within $\epsilon $ expansion near
four dimensional space-time\cite{zerf3}. Though the cubic term is relevant at leading
order,\ in the following, we will show it is irrelevant in the one-loop
corrections. This leaves the concept of FIQCP in the fermion
systems\cite{yao1,Scherer1,classen1}.

Correspondingly, the semimetal-QAH quantum criticality is governed by the
Lagrangian
\begin{equation}
\mathcal{L}_{\text{cIQAH}}=\bar{\Psi}_{i}i\gamma ^{\mu }\partial _{\mu }\Psi
_{i}+g_{\phi }\bar{\Psi}\phi \Psi +\mathcal{L}_{\phi },
\end{equation}%
\begin{equation}
\mathcal{L}_{\phi }=\frac{1}{2}(\partial _{\mu }\phi )^{2}-\frac{1}{2}%
m_{\phi }^{2}\phi ^{2}-\lambda _{\phi }\phi ^{4},
\end{equation}
We have set the boson and fermion
velocities equally to preserve the Lorentz symmetry, $v_{F}=v_{B}=1$ , which is reasonable since the the
Lorentz invariance has been argued to emergent naturally near the critical point and the velocity difference between boson and fermion is always
irrelevant in Yukawa theories\cite{janssen3,roy1}. Furthermore, for the enlarged Ising$\times $%
Ising criticality, the Landau-Ginzburg action can be written as $S=\int
d^{D}x(\mathcal{L}_{\text{cICDW-QAH}}+\mathcal{L} _{\chi \phi })$, where $%
\mathcal{L}_{\chi \phi }=\lambda _{\chi \phi }\chi ^{2}\phi ^{2}$ describes
the interaction between two Ising fields with strength $\lambda _{\chi \phi
} $.

\section{CDW Ising-criticality}\label{sec3}

In this section, we study the critical properties of CDW Ising criticality
within field-theoretic RG in $D=4-\epsilon $ space-time dimensions and the
modified minimal subtraction ($\overline{MS}$). Our starting point is the
renormalized Lagrangian%
\begin{eqnarray}
\mathcal{L}_{\text{cICDW}}^{R} &=&Z_{\Psi }\bar{\Psi}_{i}i\gamma ^{\mu
}\partial _{\mu }\Psi _{i}+\mu ^{\epsilon /2}Z_{g_{\chi }}Z_{\Psi }\sqrt{%
Z_{\chi }}g_{\chi }\bar{\Psi}\chi \Psi  \notag \\
&&+\frac{1}{2}Z_{\chi }(\partial _{\mu }\chi )^{2}-\frac{1}{2}Z_{\chi
}Z_{m_{\chi }^{2}}m_{\chi }^{2}\chi ^{2}  \notag \\
&&-\mu ^{\epsilon }Z_{\lambda _{\chi }}Z_{\chi }^{2}\lambda _{\chi }\chi
^{4},
\end{eqnarray}%
where $\mu $ denotes an renormalization energy scale, the energy scale
dependencies arises from the introduction of dimensionless coupling
constants $\lambda _{\chi }^{0}=\mu ^{\epsilon }\lambda _{\chi }$, $g_{\chi
}^{0}=g_{\chi }\mu ^{\epsilon /2}$, $m_{\chi 0}^{2}=\mu ^{2}m_{\chi }^{2}$
and the superscript $0$ means the bare quantities. We defined the
renormalized fields in terms of $\Psi _{i}^{0}=\sqrt{Z_{\Psi }}\Psi _{i}$, $%
\chi =\sqrt{Z_{\chi }}\chi ^{0}$. To determined the CDW Ising critical
behvior, the renormalization constants $Z_{\Psi }$, $Z_{\chi }$, $Z_{g_{\chi
}}$, $Z_{m_{\chi }^{2}}$, $Z_{\lambda _{\chi }}$ are perturbatively
calculated up to one-loop order, the technical details of which can be found
in Ref.~\cite{zhou1}.

\subsection{Beta functions and critical exponents}

The beta function for the coupling constants are defined as the logarithmic
derivatives with respect to the energy scale,
\begin{equation}
\beta (g_{\chi })=\frac{dg_{\chi }}{d\ln \mu },\beta (\lambda _{\chi })=%
\frac{d\lambda _{\chi }}{d\ln \mu }.
\end{equation}%
In terms of the renormalization constants, the beta functions can be represented as
\begin{equation}
\beta (g_{\chi })=(-\epsilon /2-\gamma _{g_{\chi }})g_{\chi },\beta (\lambda
_{\chi })=(-\epsilon -\gamma _{\lambda _{\chi }})\lambda _{\chi },
\end{equation}%
where $\gamma _{X}$ is defined as $\gamma _{X}=d\ln Z_{X}/d\ln \mu $ for $
X=g_{\chi }$, $\lambda _{\chi }$. We remind that the expression for the beta
fuctions differ from the previous publications\cite{zerf2,zerf3}, the difference arises
from the definition of the renormalized coupling constants. Rescaling the
coupling constants according to $\lambda _{\chi }/(8\pi ^{2})\rightarrow
\lambda _{\chi }$, $g_{\chi }^{2}/(8\pi ^{2})\rightarrow g_{\chi }^{2}$, the
beta functions are given by
\begin{eqnarray}
\beta (g_{\chi }^{2}) &=&-\epsilon g_{\chi }^{2}+(2N_{f}+3)g_{\chi }^{4}, \\
\beta (\lambda _{\chi }) &=&-\epsilon \lambda _{\chi }+4N_{f}\lambda _{\chi
}g_{\chi }^{2}+36\lambda _{\chi }^{2}-N_{f}g_{\chi }^{4}.
\end{eqnarray}%
The one-loop beta functions above agree with those in the previous
publications\cite{zerf1,mihai1}. In the limit $g_{\chi }^{2}=0$, our expressions reduce to
scalar $\phi ^{4}$ theory with $Z_{2}$ or Ising symmetry.

When the system is tuned to criticality with $m_{\chi }^{2}=0$, the
simultaneous zeros of the set of beta functions give the fixed-point which
denoeted by $(g_{\chi \ast }^{2},\lambda _{\chi \ast })$. At one-loop order,
the beta functions admit four fixed-points: the Gaussian fixed-point $(0,0)$%
, the bosonic Wilson-Fisher fixed-point $(0,\epsilon /36)$, and a pair of
Ising GNY fixed-point
\begin{equation}
(g_{\chi \ast }^{2},\lambda _{\chi \ast })_{\pm }=\left( \frac{1}{2N_{f}+3}%
\epsilon ,\frac{-(2N_{f}-3)\pm W}{72(2N_{f}+3)}\epsilon \right) ,
\end{equation}%
defining $W=(4N_{f}^{2}+132N_{f}+9)^{1/2}$. Among these fixed-points, the
infrared stable fixed-point is given by the positive one.

In addition to the beta function for the coupling constants, the beta
function for the scalar mass squared is given by
\begin{equation}
\beta (m_{\chi }^{2})=\frac{dm_{\chi }^{2}}{d\ln \mu }=-(2+\gamma _{m_{\chi
}^{2}})m_{\chi }^{2},
\end{equation}%
where $\gamma _{m_{\chi }^{2}}=d\ln Z_{m_{\chi }^{2}}/d\ln \mu $ is the
anomalous dimension for mass squared. When the system is tuned to
criticality $(m_{\chi }^{2}=0)$, the inverse correlation length exponent $%
\nu ^{-1}$ is related to the mass squared anomalous dimension by
\begin{equation}
\nu ^{-1}=2+\gamma _{m_{\chi }^{2}}(g_{\chi \ast }^{2},\lambda _{\chi \ast
}).
\end{equation}%
At one-loop order, we find $\gamma _{m_{\chi }^{2}}=-12\lambda -2N_{f}g^{2}$%
, evaluating at the criticality provides $\nu ^{-1}=2-0.8347\epsilon $ for $%
N_{f}=1$, and $\nu ^{-1}=2-0.9524\epsilon $ for $N_{f}=2$. These results
have also been calculated up to three- and four- loop in previous
literatures\cite{zerf1,mihai1}. At the QCP, it is found empirically that the pair correlation
function of the order-parameter takes the form
\begin{equation}\label{power}
\langle O_{\text{CDW}}(r)O_{\text{CDW}}(r^{\prime })\rangle \sim \frac{1}{%
|r-r^{\prime }|^{D-2+\eta _{\chi }}},
\end{equation}%
where $\eta _{\chi }$ is the order-parameter anomalous dimension
characterizing the long-range power-law decay of the pair correlation
function. In the framework of the field theory, the order-parameter
anomalous dimensions is determined by
\begin{equation}
\eta _{\chi }=\frac{1}{Z_{\chi }}\frac{dZ_{\chi }}{d\ln \mu }
\end{equation}%
and the value is evaluated at criticality. We find the one-loop result $\eta
_{\chi }=2N_{f}g_{\chi }^{2}$, the evaluation at the criticality provides $%
\eta _{\chi }=0.5714\epsilon $ for $N_{f}=2$, which agrees exactly with
Ref.~\cite{zerf1} at the corresponding order.

\subsection{The nature of Ising-order transition}

\begin{figure}[tbp]
\centering
\scalebox{0.32}{\includegraphics*{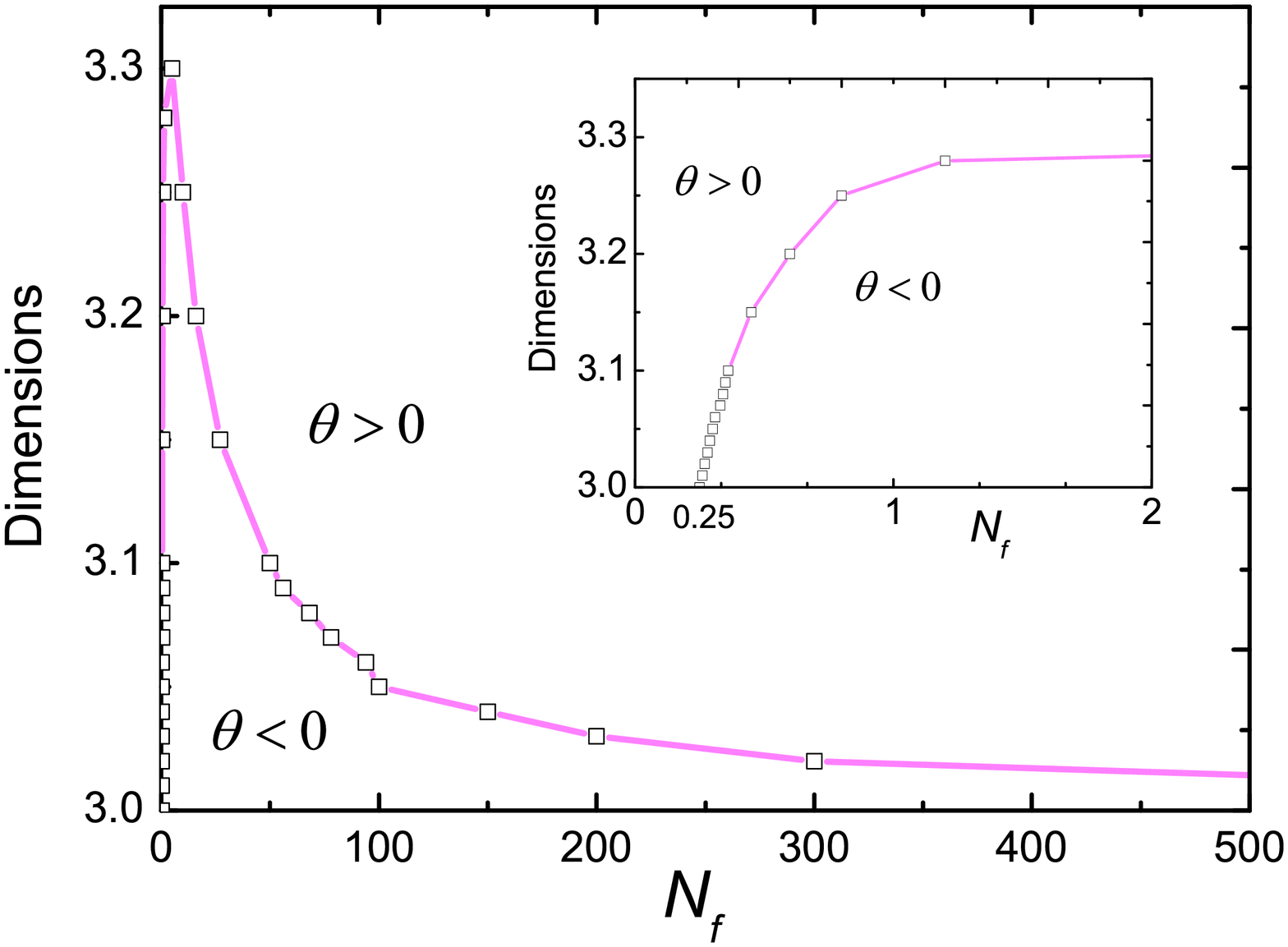}}
\caption{(Color online) Result for the range where $\protect\theta<0$ for
varying space-time dimensions and the flavors of four-component Dirac
fermion $N_f$. In this range, the cubic term is irrelevant and the phase
transition is continuous. The inset shows the detail of $N_f\in[0,2]$ and
there is a critical value $N^c_f=0.25$ for $D=3$.}
\label{fig1}
\end{figure}

So far we have considered the fixed-point without the cubic term. At
one-loop order, the cubic term contributes to the renormalization of bosonic
self-energy and cubic vertex. If the cubic term is relevant when approaching
the critical point in $4-\epsilon $ dimensions, then the transition should
be first-order. To confirm the nature of Ising transition on the honeycomb
lattice, we write down the renormalized cubic Lagrangian
\begin{equation}
\mathcal{L}_{\text{cub.}}^{R}=\mu ^{3-D/2}bZ_{b}Z_{\chi }(\chi ^{3}+\chi
^{\ast 3}),
\end{equation}
where we have introduced the renormalization constant $Z_{b}$ such that $%
b_{0}=Z_{b}b$. Similarly, the beta function for $b$ is given by
\begin{equation}
\beta (b)=[-(3-D/2)-\gamma _{b}]b,
\end{equation}
with $\gamma _{b}=d\ln Z_{b}/d\ln \mu $. Explicitly, the cubic term is
relevant (infrared) for $D\rightarrow 4$ without loop corrections. Using the
$\overline{MS}$ scheme and evaluating the one-loop correction, we find the
one-loop renormalization constant
\begin{equation}
Z_{b}=1+(36\lambda _{\chi }+3N_{f}g_{\chi }^{2})/\epsilon ,
\end{equation}
which implies the beta function
\begin{equation}
\beta (b)=[-(3-D/2)+36\lambda _{\chi }+3N_{f}g_{\chi }^{2}]b.
\end{equation}
The negative slope of the beta function $\beta (b)$ evaluated at the
fixed-point determines the relevance or irrelevance of the cubic term when
flowing toward the critical point, which gives
\begin{equation}
\theta =(3-D/2)-36\lambda _{\chi \ast }-3N_{f}g_{\chi \ast }^{2}.
\end{equation}
$\theta >0$ corresponds to a relevant and $\theta <0$ corresponds to an
irrelevant cubic coupling. At the infrared stable Ising GNY fixed-point, we
find
\begin{equation}
\theta =(3-\frac{D}{2})-\frac{4N_{f}+3+s}{2(2N_{f}+3)}\epsilon .
\end{equation}
The numerical irrelevant range is displayed in Fig.~\ref{fig1}, the dimensions $D\in
\lbrack 3,3.34]$ allows an irrelevant cubic coupling. In the $\theta >0$
range, cubic coupling is relevant, which render a first-order transition.
Instead, in the $\theta <0$ range, cubic coupling is irrelevant and we
expect a second-order transition. We also determine the second-order range
in $D=2+1$ dimensions, the one-loop RG calculations found a critical fermion
flavor number $N_{f}^{c}=1/4$. Above $N_{f}^{c}$, $\theta $ is negative so
that a continuous critical point takes place for $N_{f}\geq N_{f}^{c} $.

\begin{table}[tbp]
\caption{Numerical values of critical index for different $N_f$ in three
dimensions space-time ($D=3$), these indices determine critical behavior for
$D<3.34$.}
\label{tab1}
\begin{tabular}{llll}
\hline\hline
$N_f$\qquad\qquad\qquad\qquad & $y_1$\qquad\qquad\qquad\qquad\qquad & $y_2$
\qquad\qquad\qquad & $y_3 $ \\ \hline
0.1 & 0.2319 & -1 & -1.4737 \\
0.2 & 0.0642 & -1 & -1.7539 \\
0.25 & $1.4\times10^{-16}$ & -1 & -1.8571 \\
0.26 & -0.0117 & -1 & -1.8757 \\
0.3 & -0.0552 & -1 & -1.9437 \\
1 & -0.4042 & -1 & -1.9437 \\
10 & -0.3387 & -1 & -1.8079 \\
100 & -0.0607 & -1 & -1.1363 \\
$10^5$ & $-0.67\times10^{-4}$ & -1 & -1.0000 \\ \hline\hline
\end{tabular}
\end{table}

We have also calculated the stability matrix at criticality in three
dimensions space-time. From the beta functions, the stability matrix is
given by the linearization of flow equations at the fixed-point on the
hypersurfaces of coupling constants,
\begin{equation}
\beta (X_{i})=\mathcal{B}_{i,j}(X_{j}-X_{j}^{\ast }),
\end{equation}%
where $\mathcal{B}_{i,j}=\partial \beta _{i}/\partial
X_{j}|_{X_{j}=X_{j}^{\ast }}$ and $-\mathcal{B}_{i,j}$ is termed stability
matrix. The eigenvalues of $-\mathcal{B}_{i,j}$ define the critical
exponents which are universal at the putative continuous critical point.
More explicitly, one has
\begin{equation}
\frac{dX_{i}}{d\ln s}=-\mathcal{B}_{i,j}(X_{j}-X_{j}^{\ast }),
\end{equation}%
the renormalization scaling factor $s$ is accompanied by the relation $%
\mathcal{F}(X_{i})\sim s^{-D}\mathcal{F}(s^{y_{i}}X_{i})$, where $\mathcal{F}
$ is an universal scaling function. Choosing $\vec{X}=(g_{\chi}^{2},\lambda
_{\chi },b)$, we have three eigenvalues which are ordered as $%
y_{1}>y_{2}>y_{3}$. A second-order critical point requires $y_1<0$. The
critical index $y_i>0$ implies $X_{i}$ is a relevant variable, this
corresponds to repelling flow. By contrast, $y_{i}<0$ implies $X_{i}$ is an
irrelevant variable, this corresponds to attractive flow. Our calculations
of the critical index in $D=3$ dimensions are listed in Table \ref{tab1}, we find the
critical index $y_{1}$ change sign at $N_{f}^{c}=0.25$, which gives a
consistent check on the critical fermion flavor number. Consequently, the
two-dimensional honeycomb lattice with $N_{f}=2$ may display a continuous
phase transition with universal critical behavior.

\section{Ising$\times $Ising criticality}\label{sec4}
\begin{figure}[tbp]
\centering
\scalebox{0.85}{\includegraphics*{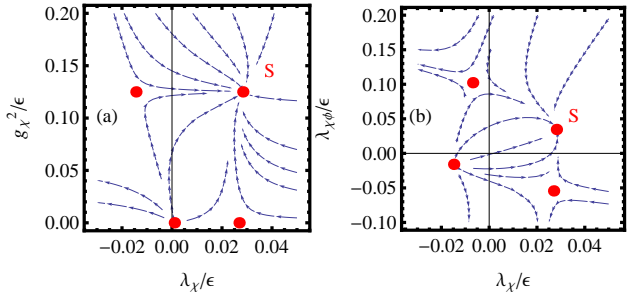}}
\caption{(Color online) The fixed points and the RG flow within the space
spanned by $g_{\protect\chi }^{2}$, $\protect\lambda _{\protect\chi }$ and $
\protect\lambda _{\protect\chi \protect\phi }$ for $N_{f}=1$. (a) $\protect
\lambda _{\protect\chi }$-$g_{\protect\chi }^{2}$ plane for $\protect\lambda
_{\protect\chi \protect\phi }=0.0346\protect\epsilon $. (b) $\protect\lambda
_{\protect\chi }$-$\protect\lambda _{\protect\chi \protect\phi }$ plane for $
g_{\protect\chi }=0.125\protect\epsilon $. The fixed point denoted by $S$ (red) is
an infrared stable fixed-point.}
\label{FS}
\end{figure}

The previous section focus only on the Ising criticality, we now turn to the
critical behavior with enlarged Ising$\times $Ising fluctuations. The model
under consideration is given by $\mathcal{L}_{\text{II}}=\mathcal{L}_{\text{
cICDW-QAH}}+\mathcal{L}_{\chi \phi }$, and the corresponding renormalized Lagrangian is given by
\begin{align}
\mathcal{L}_{\text{II}}^{R}& =Z_{\Psi }\bar{\Psi}_{i}i\gamma ^{\mu }\partial
_{\mu }\Psi _{i}+\mu ^{\epsilon }Z_{\lambda _{\chi \phi }}Z_{\chi }Z_{\phi
}\lambda _{\chi \phi }\chi ^{2}\phi ^{2}  \notag \\
& +\frac{1}{2}Z_{\chi }(\partial _{\mu }\chi )^{2}-\frac{1}{2}Z_{\chi
}Z_{m_{\chi }^{2}}m_{\chi }^{2}\chi ^{2}-\mu ^{\epsilon }Z_{\lambda _{\chi
}}Z_{\chi }^{2}\lambda _{\chi }\chi ^{4}  \notag \\
& +\frac{1}{2}Z_{\phi }(\partial _{\mu }\phi )^{2}-\frac{1}{2}Z_{\phi
}Z_{m_{\phi }^{2}}m_{\phi }^{2}\phi ^{2}-\mu ^{\epsilon }Z_{\lambda _{\phi
}}Z_{\phi }^{2}\lambda _{\phi }\phi ^{4}  \notag \\
& +\mu ^{\frac{\epsilon }{2}}Z_{g_{\chi }}Z_{\Psi }Z_{\chi }^{1/2}g_{\chi }%
\bar{\Psi}_{i}\chi \Psi _{i}+\mu ^{\frac{\epsilon }{2}}Z_{g_{\phi }}Z_{\Psi
}Z_{\phi }^{1/2}g_{\phi }\bar{\Psi}_{i}\phi \Psi _{i}.
\end{align}
Similar models have been used to discuss the coexisting orders and Mott mutilcriticality in Dirac systmes, see Refs.~\cite{torres2,classen3}. As defined is Sec.~\ref{sec3}, the $Z_{i}$ renormalization factors. We
have also introduced the interaction between the CDW dynamical fluctuating
and QAH dynamical fluctuating, which is given by $\mathcal{L}_{\chi\phi}$. Evaluating these renormalization factors at
one-loop order, the beta functions of the rescaled coupling constants are
given by the following differential equations

\begin{align}
\beta (g_{\chi }^{2})& =-\epsilon g_{\chi }^{2}+(2N_{f}+3)g_{\chi
}^{4}+3g_{\phi }^{2}g_{\chi }^{2},  \label{b1} \\
\beta (g_{\phi }^{2})& =-\epsilon g_{\phi }^{2}+(2N_{f}+3)g_{\phi
}^{4}+3g_{\chi }^{2}g_{\phi }^{2},  \label{b2} \\
\beta (\lambda _{\chi })& =-\epsilon \lambda _{\chi }+4N_{f}\lambda _{\chi
}g_{\chi }^{2}+36\lambda _{\chi }^{2}-N_{f}g_{\chi }^{4}+\lambda _{\chi \phi
}^{2}  \label{b3} \\
\beta (\lambda _{\phi })& =-\epsilon \lambda _{\phi }+4N_{f}\lambda _{\phi
}g_{\phi }^{2}+36\lambda _{\phi }^{2}-N_{f}g_{\phi }^{4}+\lambda _{\chi \phi
}^{2},  \label{b4} \\
\beta (\lambda _{\chi \phi })& =-\epsilon \lambda _{\chi \phi
}+2N_{f}\lambda _{\chi \phi }g_{\chi }^{2}+2N_{f}\lambda _{\chi \phi
}g_{\phi }^{2}+8\lambda _{\chi \phi }^{2}  \notag \\
& +12\lambda _{\chi }\lambda _{\chi \phi }+12\lambda _{\phi }\lambda _{\chi
\phi }-N_{f}g_{\phi }^{2}g_{\chi }^{2},  \label{b5}
\end{align}%
To confirm the fixed-point on the critical hypersurface denoted by $%
X_{i}^{\ast }=(g_{\chi }^{2},g_{\phi }^{2},\lambda _{\chi },\lambda _{\phi
},\lambda _{\chi \phi })^{\ast }$, we look for the solution for the simultaneous zero of these beta functions, $\beta
(X_{i}^{\ast })=0$. Eqs.(\ref{b1}) and (\ref{b2}) admit four solutions, $
A_{1}$: $(g_{\chi }^{2},g_{\phi }^{2})=(0,0)$, $A_{2}$: $[\epsilon
/(2N_{f}+3)$,$0]$, $A_{3}$: $[0$, $\epsilon /(2N_{f}+3)]$ and
\begin{equation}
A_{4}:g_{\chi }^{2}=\frac{\epsilon }{2N_{f}+6},\quad g_{\phi }^{2}=\frac{%
\epsilon }{2N_{f}+6}\text{.}
\end{equation}
Among these solutions, only $A_{4}$ corresponds to a stable fixed-point\cite{zerf1}. The
equations for $\lambda _{\chi }$ and $\lambda _{\phi }$ are symmetric, so
they enjoy the same value at criticality. We have solved the fixed-point
numerically, for instance for $N_{f}=1$, Eqs.(\ref{b1}) to (\ref{b5}) admit
an infrared stable fixed point:
\begin{equation}
X^{\ast }=(0.125\epsilon ,0.0281\epsilon ,0.0281\epsilon ,0.0346\epsilon
,0.0346\epsilon ),
\end{equation}%
at which the critical behavior is universal. The stable infrared fixed-point
and the RG flow spanned by $g_{\chi }^{2}$, $\lambda _{\chi }$ and $\lambda
_{\chi \phi }$ are illustrated in Fig.~\ref{FS}. In the presence of enlarged
Ising$\times $Ising fluctuations, we observe that the enlarged fluctuation
brings the system to a dual GNY fixed-point denoted by $S$.
\subsection{Critical exponents}

We now turn to the computation of critical exponents. At one-loop order, the
fermion field renormalization has additional contribution compared with the
Ising criticality, it is easily calculated the field renormalization
coefficient (see details in Ref.~\cite{zhou1}):
\begin{equation}
Z_{\psi }=1-(g_{\chi }^{2}+g_{\phi }^{2})/(2\epsilon ),
\end{equation}%
which yields the fermion anomalous dimensions $\eta _{\psi }=(g_{\chi
}^{2}+g_{\phi }^{2})/2$ such that $2\Delta _{\psi }=D-1+\eta _{\psi }$,
where $\Delta _{\psi }$ is the scaling dimensions for Dirac fermions. As stated in
Sec.~\ref{sec3}A, the boson anomalous dimensions are given by $\eta _{\chi
}=2N_{f}g_{\chi }^{2}$ and $\eta _{\phi }=2N_{f}g_{\phi }^{2}$, respectively.
The inverse correlation length exponent characterizes the divergence of
correlation length as the mass squared is tuned to zero or the transition is
approached. At the critical point, we find the inverse correlation length
exponent, at one-loop order, is given by
\begin{equation}
1/\nu =2-12\lambda _{\chi \ast }-2N_{f}g_{\chi \ast }^{2}-\lambda _{\chi
\phi \ast }.
\end{equation}
Interestingly, the GNY model with Ising$\times $Ising
criticality also supports the emergent supersymmetry (SUSY) scenario \cite{yao5,grover1}. For $N_{f}=1/2$
, the quantitative estimates of the critical exponents finds
\begin{eqnarray}
1/\nu &=&2-0.5217\epsilon , \\
\eta &=&\epsilon /7,
\end{eqnarray}%
with $\eta =\eta _{\chi }=\eta _{\phi }$. We point also that, owing to the
existence of strongly relevant mixed term between two different Ising
fluctuating fields, the supersymmetry scaling relation $1/\nu =(D-\eta )/2$
in the chiral Ising GNY model not holds exactly at criticality\cite{zerf1}. In general
case for $N_{f}\geq 1/4$, the critical exponents define a new universality
class termed chiral Ising$\times$Ising universality class. To obtain the
estimate for the critical exponents at the physical dimensions $\epsilon =1$%
, we employ simple Pade approximant\cite{ihrig1}. At one-loop order, the Pade
approximant provides only [0/1] extrapolation for the exponents, our
estimates for different $N_{f}$ are listed in the Table.~\ref{tab2}.
For $N_{f}=1$, the theory describes the quantum criticality of spinless electrons from semimetallic state to
insulating state which break sublattice or $Z_2$ symmetry, i.e., CDW phase. The $N_{f}=2$ GNY model describes the similar transition of spinful fermions on the honeycomb lattice.

\begin{table}[tbp]
\caption{ Critical exponents for the chiral Ising$\times$Ising
universality class in $D=3$ for varying flavors of Dirac fermion $N_{f}$: inverse
correlation length exponent $1/\nu$, boson anomalous dimension $\eta_{\chi}$, $\eta_{\phi}$, and fermions anomalous dimension $\eta_{\psi}$.
We provide Pade estimate for the correction length exponents with [0/1]
extrapolation.}
\label{tab2}
\begin{tabular}{llll}
\hline\hline
$N_f$\qquad\qquad\qquad & $\nu^{-1}_{[0/1]}$\qquad\qquad\qquad\qquad & $%
\eta=\eta_{\chi}=\eta_{\phi}$ \qquad\qquad & $\eta_{\psi} $ \\ \hline
1/4 & 1.6318 & 0.0769 & 0.1538 \\
1/2 & 1.5863 & 0.1478 & 0.1428 \\
1 & 1.5254 & 0.25 & 0.125 \\
2 & 1.4589 & 0.4 & 0.1 \\ \hline\hline
\end{tabular}
\end{table}

\subsection{Scaling dimensions of fermion bilinears}

Apart from the order-parameter anomalous dimensions, the pair correlation
function of fermion bilinear also develops universal long-range power-law
decay at criticality. It's very interesting to ask for the behavior of the
bilinear correlation at criticality. In general, the microscopic
order-parameter on an underlying lattice model can be identified with the
bilinears obtained in the continuum limit, i.e., valence-bond-solid order or
Neel order\cite{hermele1,ghaemi1}. So, the observable value of fermion bilinears is accessible to
quantum Monte Carlo simulations\cite{meng2}. The fermion bilinears are gauge-invariant
while the fermion fields anomalous dimensions are not in a gauge theory such
as QED theory, which allows us to calculate the scaling dimensions of the fermion
bilinear.

Following Ref.~\cite{zerf2}, we add an infinitesimally weak fermion bilinear $m_{0}\bar{
\Psi}_{0}M\Psi _{0}$ in the bare Lagrangian, then the renormalized quantity is
given by $Z_{m}Z_{\Psi }m_{R}\bar{\Psi}M\Psi $. The beta function for
the weak mass reads
\begin{equation}
\beta (m)=\frac{dm}{d\ln \mu }=-(1+\gamma _{m})m,
\end{equation}%
where $\gamma _{m}=d\ln Z_{m}/d\ln \mu $ is the anomalous dimensions for $m$%
. The scaling dimensions of the bilinear is then given by
\begin{equation}
\Delta _{\langle \bar{\Psi}M\Psi \rangle }=D-1-\gamma _{m}(g_{\chi \ast
}^{2},\lambda _{\chi \ast }).
\end{equation}%
It's straightforward to calculate the scaling dimensions\ of QAH fermion
bilinear in the Ising criticality, we find
\begin{equation}
\Delta _{\langle \bar{\Psi}\gamma ^{35}\Psi \rangle }=3-\frac{4N_{f}+3}{%
4N_{f}+6}\epsilon +O(\epsilon ^{2}),
\end{equation}
this result is also concide with the scaling dimensions of the flavor
singlet and adjoint fermion bilinears calculated in the chiral Ising GNY
model\cite{zerf3}. Since the anticommutating relation $[\gamma ^{35},\gamma ^{\mu }]=0$%
, the CDW and QAH fermion bilinear are expected to have the same scaling
dimensions at one-loop order $\Delta _{\langle \bar{\Psi}\gamma ^{35}\Psi
\rangle }=\Delta _{\langle \bar{\Psi}\Psi \rangle }$. Similally, the CDW and
QAH bilinear at criticality show power-law decay as Eq.~(\ref{power}), and the scaling
dimensions at the Ising$\times $Ising criticality is given by
\begin{equation}
\Delta _{\langle \bar{\Psi}\gamma ^{35}\Psi \rangle }=\Delta _{\langle \bar{
\Psi}\Psi \rangle }=3-\frac{2N_{f}+3}{2N_{f}+6}\epsilon +O(\epsilon ^{2}),
\end{equation}%
which is apparently larger than those in the Ising criticality for
relatively small $N_{f}$. Therefore, the QAH bilinear correlations at the
Ising$\times$Ising criticality decay faster than it at the Ising criticality.

Further, the bilinears $\langle \bar{\Psi}\gamma ^{3}\Psi \rangle $ and $
\langle \bar{\Psi}\gamma ^{5}\Psi \rangle $ are of interests, they
correspond to Kekule VBS order parameter on the honeycomb lattice\cite{chamon1,ryu1}. The
corresponding scaling dimensions in the Ising and Ising$\times $Ising
criticality can be calculated respectively as
\begin{align}
\Delta _{\text{KVBS}}^{\text{Ising}}& =3-\frac{4N_{f}+15}{4N_{f}+12}\epsilon
+O(\epsilon ^{2}), \\
\Delta _{\text{KVBS}}^{\text{IxI}}& =3-\frac{2N_{f}+9}{2N_{f}+6}\epsilon
+O(\epsilon ^{2}),
\end{align}%
with $\Delta _{\text{KVBS}}=\Delta _{\langle \bar{\Psi}\gamma ^{3}\Psi
\rangle }=\Delta _{\langle \bar{\Psi}\gamma ^{5}\Psi \rangle }$. An
important observation from this result is that the Kekule VBS correlation has
been enhanced tremendously by the Ising$\times $Ising fluctuations as the bilinear
scaling dimensions is decreased.

By now, we have concentrated on the Ising$\times $Ising criticality of
spinless electrons. The spinful electrons on the graphene lattice are
believed to undergo a metal-insulator phase transition for repulsive
interactions. Using the eight-component spinor $(\Psi _{i\uparrow },\Psi
_{i\downarrow })$, the gamma matrices in the kinetic part can be written as
[see Eq.(\ref{gamma})] $\Gamma ^{0}=s^{0}\otimes \tau ^{0}\otimes \sigma ^{3}$, $\Gamma
^{1}=s^{0}\otimes \tau ^{0}\otimes i\sigma ^{1}$, $\Gamma ^{2}=s^{0}\otimes
\tau ^{3}\otimes i\sigma ^{2}$, where the Pauli matrix $s^{i}$ acts as the
real spin degrees of freedom. In addition to the bilinears already
encountered in the spinless case, there exist 12 bilinears for different
microscopic order parameter\cite{chamon1}. The typical example are Neel order, quantum
spin Hall effect (QSHE) and spin-dependent Kekule VBS. We denote the
bilinears by $\langle \bar{\Psi}M_{O}\Psi \rangle $, for example,
$M_{O}=\vec{s}\otimes
\tau ^{0}\otimes \sigma ^{0}$ corresponds to the spin-density-wave or
z-direction Neel order
\begin{equation}
O_{\text{Neel}}^{z}=(-1)^{\text{i}}\langle c_{i\uparrow }^{\dag
}c_{i\uparrow }-c_{i\downarrow }^{\dag }c_{i\downarrow }\rangle .
\end{equation}
$M_{O}=\vec{s}\otimes \tau ^{3}\otimes i\sigma ^{0}$ corresponds to the
QSHE. Finally, both $M_{O}=\vec{s}\otimes \tau ^{1}\otimes i\sigma ^{2}$ and
$M_{O}=$ $\vec{s}\otimes \tau ^{2}\otimes i\sigma ^{2}$ correspond to the
Kekule VBS\cite{chamon1}, their combination
\begin{equation}
O_{\text{KVBS}}=\Gamma ^{0}(\vec{s}\otimes \tau ^{1}\otimes i\sigma ^{2})+%
\text{i}\Gamma ^{0}(\vec{s}\otimes \tau ^{2}\otimes i\sigma ^{2})
\end{equation}
controls the dimerization pattern. For these bilinears, we find only the
Kekule VBS scaling dimensions at the Ising$\times $Ising critical point is
smaller that its value at the Ising critical point. Thus, the kekule VBS
correlations are enhanced tremendously by the Ising$\times $Ising fluctuations while
other correlations are suppressed. In general, it is worth pointing out that
the bilinear correlations are enhanced if the bilinear matrices $M_{O}$ and the gamma matrices $\Gamma ^{\mu }$ are anticommutative, namely $\{\Gamma^{\mu },M_{O}\}=0$.
In summary, the crucial ingredient for the enhancement is the anticommuting nature between the corresponding fermion bilinear matrix and the Dirac gamma matrices.

\section{Conclusions}\label{sec5}

In this paper, with the help of one-loop perturbative RG analysis in $d=4-\epsilon$,
we have studied the fermionic quantum criticality with enlarged Ising$\times$Ising fluctuation
in the two dimensional honeycomb materials.
To get an understanding of whether the semimetal-insulator transition is weak first-order or second-order
transition, we include a cubic term of the order-parameter in the theory and study its fate as the infrared stable fixed-point is approached.
The semimetal-CDW transition is modeled in terms of an Ising GNY theory with a generalized flavors of Dirac fermions $N_f$,
we find the cubic term is always irrelevant if $N_f$ fulfills $N_f\geq1/4$ in three space-time dimensions. The irrelevance implies that the extra fluctuations from fermions change of the nature of transition and render it continuous\cite{yao2,yin1,classen1}.
We also calculate the complete second-order regime for varying space-time and $N_f$, as shown in Fig.~\ref{fig1}.

Moreover, the tricritical point for the semimetal-transition that breaks Ising$\times$Ising symmetry is investigated. Using $\epsilon$ expansion,  we have calculated the critical exponents for the Ising$\times$Ising universality class, including inverse correlation length exponent $1/\nu$, boson anomalous dimension $\eta_{\chi}$, $\eta_{\phi}$, and fermions anomalous dimension $\eta_{\psi}$. The exponents for different
value of $N_f$ are shown in Table \ref{tab2}.
Further, the $\epsilon$ expansion has been used to calculated the scaling dimensions for the fermion bilinear on the honeycomb lattice.
In particular, we observe that the scaling dimensions for the Kekule valence-bond-solid at Ising$\times$Ising criticality is smaller than the value at Ising criticality. This means that the Kekule valence-bond-solid is enhanced tremendously by the enlarged Ising$\times$Ising fluctuations. The crucial ingredient for the enhancement is the anticommuting nature between the corresponding fermion bilinear matrix and the Dirac gamma matrices. We hope such kind of tremendous enhancement will shed light on the mutil-criticality of competing orders in complex many-body systems and even the transition in the high-temperature superconductors.

\begin{acknowledgments}
This work has been supported by the NSFC under No.11647111, No.11674062,
No.11974053, we also acknowledge the support by the Scientific Research Foundation of Guizhou
University under No.20175788 (J.Zhou).
\end{acknowledgments}

\end{document}